\documentclass[12pt]{article}
\def\beq {\begin{eqnarray}}
\def\eeq {\end{eqnarray}}
\def\be {\begin{equation}}
\def\ee {\end{equation}}
\def \12 {{\textstyle{1\over 2}}}

\def\Tr {{\rm Tr}}
\def\dag {\dagger}
\def\del {\partial}

\def\a {\alpha}
\def\b {\beta}

\def\d {\delta}
\def\D {\Delta}
\def\bz {\bar{z}}

\def\ra {\rangle}
\def\la {\langle}
\def\Tr {{\rm Tr}}
\def\bJ {{\bar J}}
\def\bxi{{\bar \xi}}

\def\bz {{\bar z}}
\def\bphi{\bar{\phi}}

\def \vA {{\vec A}}
\def \vx {{\vec x}}

\begin{document}

\begin{titlepage}
\null\vspace{-62pt}

\pagestyle{empty}
\begin{center}
\rightline{CCNY-HEP-01/03}

\vspace{1.0truein} {\large \bf  Chern-Simons matrix model:
coherent states} \\ \vspace{.2in} {\large\bf  and relation to
Laughlin wavefunctions} \\ \vspace{.5in}  { Dimitra Karabali
$^{a,c}$ and  B. Sakita $^{b,c}$ \footnote{{\it e-mail addresses:}
karabali@alpha.lehman.cuny.edu, sakita@scisun.sci.ccny.cuny.edu}
}\\ \vspace{.3in}  {\it $^a$ Department of Physics and Astronomy,
Lehman College of the CUNY\\ Bronx, NY 10468}\\ \vspace {.1in}
{\it $^b$ Physics Department, City College of the CUNY\\  New
York, NY 10031}\\ \vspace {.1in} {\it $^c$ The Graduate School and
University Center, CUNY\\ New York, NY 10016}\\

\vspace{0.5in}
\end{center}
\vspace{0.5in}

\centerline{\bf Abstract}

Using a coherent state representation we derive many-body
probability distributions and wavefunctions for the Chern-Simons matrix
model proposed by Polychronakos and compare them to the Laughlin
ones. We analyze two different coherent state representations,
corresponding to different choices for electron coordinate bases. In both
cases we find that the resulting probability distributions do not quite
agree with the Laughlin ones. There is agreement on the long distance
behavior, but the short distance behavior is different.

\baselineskip=18pt

\end{titlepage}

\hoffset=0in
\newpage
\pagestyle{plain}
\setcounter{page}{2}
\newpage

\noindent
{\bf 1. Introduction}

There has recently appeared an interesting connection between quantum Hall effect and
noncommutative field theory. In particular Susskind proposed in \cite{susskind}, that
the Laughlin states at filling fractions $\nu ={1 \over {2p+1}}$ for a system of an
infinite number of electrons confined in the lowest Landau level (LLL) can be described
by a noncommutative $U(1)$ Chern-Simons theory. The fields of this theory are infinite
matrices which act on an infinite Hilbert space, appropriate to account for an infinite
number of electrons. In the same spirit, Polychronakos, later, proposed a finite matrix
model \cite{poly1}, a regularized version of this noncommutative Chern-Simons theory in
an effort to describe finite systems, of limited spatial extent with a finite number of
electrons. This matrix model was shown to reproduce the basic features of the quantum
Hall droplets and their corresponding excitations \cite{girvin}, such as boundary and
quasihole excitations at filling fraction $\nu ={1 \over {2p+1}}$.

In a subsequent paper \cite{states}, Hellerman and Raamsdonk, trying to make the
connection between quantum Hall effect and the noncommutative matrix model more
transparent, analyzed the states of the theory and concluded that the states of the
matrix model are in one-to-one correspondence with the Laughlin states describing QHE
at filling fraction
$\nu = {1
\over {2p+1}}$. Similar arguments were put forward for the excited states.
Although this is an interesting observation, the existence of a one-to-one
mapping between states is not enough to prove the equivalence of the two
theories.

Since the mapping in \cite{states} is somewhat formal at the level of states, in this
paper we try to go one step further and compare the two theories at the level of the
wavefunctions. This requires a notion of coordinates, which is
introduced via a coherent state representation for the states of the matrix model. We
will present two different ways of arriving at a coherent state representation, which
are not equivalent in general. In the special case of $\nu =1$ ( $\theta=0$), it turns
out that both coherent state representations of the corresponding matrix model
reproduce the $\nu=1$ Laughlin wavefunction. For $\nu ={1
\over {2p+1}}$ though, the two ways produce different probability distributions and
neither agrees with the Laughlin one. The probability distributions emerging from
the different coherent state representations have a long distance behavior
similar to the corresponding Laughlin one, but quite a different short distance
behavior.

This paper is organized as follows. In section 2 we give a brief discussion of fermions
in the lowest Landau level and the Laughlin wavefunctions. In section 3 we briefly
review the Chern-Simons finite matrix model proposed by Polychronakos. In section 4 we
present two different approaches towards a coherent state representation of the matrix
states. In section 5 we comment on our results.
\vskip .2in
\noindent
{\bf 2. Fermions in the lowest Landau level}

It is well known by now that the two-dimensional configuration space of charged
particles in the LLL is equivalent to a one-dimensional phase space and therefore
noncommutative \cite{jackiw, sakita}. To see this let us consider a charged particle
in the presence of a strong uniform magnetic field. The Lagrangian is given by
\be
L= {m \over 2} {\dot \vx}^2 + \vA \cdot \dot{\vx} - V(x_1, x_2)
\label{charge}
\ee
where $V(x_1, x_2)$ is an abritrary confining potential. For convenience we shall
consider $A_0$ to be a harmonic oscillator potential of the form
\be
V = {1 \over 2} \omega \vx ^2
\label{harmonic}\ee
The energy eigestates of this system lie on Landau levels and at the limit $m
\rightarrow 0$ (or equivalently strong $B$) the system is confined to the LLL. In this
case the kinetic energy term in (\ref{charge}) is negligible. We further use the radial
gauge $A_i =  {1 \over 2} B \epsilon_{ij} x_j$. The canonical momentum of $x_1$ is
then $p_1= Bx_2$ and as a result
\be
[x_1,~x_2] = [x_1,~ {p_1 \over B}] = {i \over B}
\label{CCR}
\ee

In the absence of a confining potential all the eigenstates at each Landau level are
degenerate. The confining potential lifts the degeneracy. In the lowest Landau level
the Hamiltonian is
\be
H = {\omega \over B} ( a^{\dagger} a + {1 \over 2})
\label{H}
\ee
where $a = {\sqrt{B \over 2}} (x_1+ix_2)$. The single particle eigenstates are
\be
|n\ra = {1 \over {\sqrt{n!}}}~~a^{\dag n} |0 \ra
\label{a}
\ee
The corresponding normalized wavefunctions are
\be
\Psi _n =\sqrt{B \over {2 \pi n!}} ~~\bz^n e ^{- |z|^2/2}
\label{sw}
\ee
where $z = \sqrt{{B \over 2}} (x_1 + i x_2)$.
These can be thought of as the coherent state representation of (\ref{a}).

Consider now the case of $N$ fermions. Since each state can be occupied by at most one
fermion, the presence of the confining potential selects a unique ground state which is
the minimum angular momentum state. This is the $\nu =1$ ground state wavefunction
\be
\Psi_1 (\vx _1, ..., \vx _N) = \prod_{i<j} (\bz_i - \bz_j) e^{-\12 \sum_{i=1}^{N}
|z_i|^2}
\label {nu1}
\ee
This corresponds to an incompressible circular droplet configuration of uniform density
$\rho = {B \over 2\pi}$. This incompressibility is crucial in explaining the
experimentally observed gap for $\nu =1$. The existence of a similar gap is less
obvious in the case of noninteger filling fractions, where the LLL is only partially
filled. In this case, it is believed that the repulsive Coulomb interactions among
electrons are important in generating strong correlations which eventually produce a
new ground state with a gap. Laughlin proposed that such a state, in the case $\nu = {
1 \over {2p+1}}$, is described by a wavefunction \cite{laughlin}
\be
\Psi_{2p+1} (\vx _1,...,\vx _N) = \prod _{i<j}^N (\bz_i -\bz_j)^{2p+1} e^{-\12
\sum_{i=1}^{N} |z_i|^2}
\label{Laughlin}
\ee
Using the connection to the one-component, two-dimensional plasma, Laughlin showed that
this corresponds to an incompressible droplet of density $\rho = {B \over {2 \pi
(2p+1)}}$. Although not exact eigenfunctions, (\ref{Laughlin}) are quite close to the
true solutions, at least numerically. They vanish quite rapidly if any two particles
approach each other, and this helps minimize the expectation value of the Coulomb
energy. Their success lies very much on their short distance behavior.

Our aim is to develop an appropriate coherent state representation for the states of
the corresponding matrix model and compare them directly to the Laughlin wavefunctions.
Before we explain this in more detail, we shall give a brief review of the matrix
Chern-Simons model proposed by Polychronakos.

\vskip .2in
\noindent
{\bf 3. Chern-Simons matrix model}

The action describing the matrix Chern-Simons model is given by \cite{poly1}
\be
S = \int dt {B \over 2} Tr \{ \epsilon _{ab} (\dot{X} _a + i [A_0,~ X_a]) X_b + 2
\theta A_0 - \omega X_a ^2\} + \Psi^{\dag}(i \dot{\Psi} - A_0 \Psi)
\label{action}
\ee
where $X_a,~a=1,2$ are $N \times N$ matrices and $\Psi$ is a complex $N$-vector that
transforms in the fundamental of the gauge group $U(N)$,
\be
X_a \rightarrow U X_a U^{-1}~~,~~~~~~~~~\Psi \rightarrow U \Psi
\label{transf}
\ee
The $A_0$ equation of motion implies the constraint
\be
G \equiv -iB [X_1,~X_2] + \Psi \Psi^{\dag} -B \theta =0
\label{constraint}
\ee
The trace of this equation gives
\be
\Psi^{\dag} \Psi = N B \theta
\label{trace}
\ee
It is interesting to note that in the absence of $\Psi$'s, the parameter $\theta$ has
to be zero. In this case the action (\ref{action}) is that of a one-dimensional
hermitian matrix model in a harmonic potential, which is known to be equivalent to $N$
one-dimensional fermions in a  confining potential \cite{brezin}.

Upon quantization the matrix elements of $X_a$ and the components of $\Psi$ become
operators, obeying the following commutation relations
\beq
\bigl[ \Psi_i,~ \Psi_j^{\dag} \bigr] & = & \delta _{ij} \nonumber  \\
\bigl[ (X_1 ) _{ij},~ (X_2 ) _{kl} \bigr] & = & {i \over B} \delta _{il} ~ \delta _{jk}
\label{CR}
\eeq

The Hamiltonian is
\be
H = \omega ( { N^2 \over 2} + \sum A_{ij}^{\dag} A_{ji})
\label{hamiltonian}
\ee
where $A = \sqrt{B \over 2} (X_1 + i X_2)$.
The system contains $N(N+1)$ oscillators coupled by the constraint (\ref{constraint}).
As explained in \cite{poly1}, upon quantization, the operator $G$ becomes the
generator of unitary rotations of both
$X_a$ and $\Psi$. The trace part (\ref{trace}) demands that $NB \theta $ being the
number operator for $\Psi$'s is quantized to an integer. The traceless part of the
constraint demands the physical states to be singlets of $SU(N)$.

Since the $A_{ij}^{\dag}$ transform in the adjoint and the $\Psi_{i}^{\dag}$
transform in the fundamental representation of $SU(N)$, a purely group theoretical
argument implies that a physical state being a singlet has to contain $N n$
$\Psi^{\dag}$'s, where $n$ is an integer. This leads to the quantization of $B \theta
=k$. We shall see later that in identifying this matrix model with a fermionic system,
$k$ has to be an even integer.

Explicit expressions for the states were written down in \cite{states}. 
(Essentially equivalent results were also obtained by Polychronakos
\cite{poly3}.) The ground
state being an
$SU(N)$ singlet with the lowest number of $A^{\dag}$'s is of the form
\be
\vert \Psi \ra = \bigl[ \epsilon ^{i_1...i_N} \Psi^{\dag}_{i_1} (\Psi^{\dag}
A^{\dag})_{i_2}...(\Psi^{\dag} A^{\dag N-1})_{i_N} \bigr] ^k \vert 0 \ra
\label{state}
\ee
where $\vert 0 \ra$ is annihilated by $A$'s and $\Psi$'s, while the excited states can
be written as
\be
\vert \Psi_{exc} \ra = \prod_{i=1}^{N-1} ( \Tr A^{\dag i})^{c_i} \bigl[ \epsilon
^{i_1...i_N}
\Psi^{\dag}_{i_1} (\Psi^{\dag} A^{\dag})_{i_2}...(\Psi^{\dag} A^{\dag N-1})_{i_N}
\bigr] ^k \vert 0 \ra
\label{excited}
\ee

\vskip .2in
\noindent
{\bf 4. Coherent state representations}

We now present a coherent state representation for the matrix states (\ref{state},
\ref{excited}). In doing this one has to make a choice of coordinates, which eventually
parametrize the phase space of the underlying one-dimensional system.

{}From the matrix model point of view, a natural choice would be to diagonalize $A$ and
interpret its eigenvalues as phase space coordinates of particles. We refer to this as the
$A$-representation. This is essentially a generalized complex random matrix model.

Another choice would be to diagonalize $X_1$ and interpret its eigenvalues as
one-dimensional coordinates. This is the $X$-representation. In this, the elements of
$X_2$ are canonically conjugate to $X_1$. Once the $X$-representation of the matrix
states (\ref{state}, \ref{excited}) is derived, we can then use the usual coherent state
representation to express the wavefunctions in terms of phase space coordinates.

Clearly in the two cases, $A$- and $X$-representation, the notion of phase space
coordinates is different; as a result we derive different expressions for the
corresponding wavefunctions.

Below we present in detail the two representations and
 compare the results to the
Laughlin wavefunctions.

\noindent
$ \underline{ A-{\rm representation}}$

 We define the coherent state $\vert Z,\phi \ra $ such
that
\beq
A_{mn} \vert Z,\phi \ra &=& Z_{mn} \vert Z,\phi \ra \nonumber \\
\Psi_n \vert Z,\phi \ra  &=& \phi_n ~~\vert Z,\phi \ra
\label{coherent}
\eeq
where $Z$ is a complex $N \times N$ matrix and $\phi$ is a complex vector. Let us
consider the matrix ground state of the form
\be
\vert 2p+1 \ra = \bigl[ \epsilon ^{i_1...i_N} \Psi^{\dag}_{i_1} (\Psi^{\dag}
A^{\dag})_{i_2}...(\Psi^{\dag} A^{\dag N-1})_{i_N} \bigr] ^{2p} \vert 0 \ra
\label{statep}
\ee
and reexpress its scalar product in terms of the coherent state wavefunctions
using the completeness relation
\be
\la 2p+1 \vert 2p+1 \ra = \int \prod_{i,j,l}^N dZ_{ij}~ dZ^{*}_{ij} ~d \phi_l ~d \bphi_l
 \la 2p+1 \vert Z, \phi \ra \la Z,
\phi \vert 2p+1 \ra
\label{inner}
\ee
Before we talk about the general case, it is interesting to demonstrate how the
coherent state representation works for $p=0$. The scalar product (\ref{inner}) can be
written as,
\be
\la 1 \vert 1 \ra = \int \prod_{i,j,l}^N dZ_{ij} dZ^{*}_{ij} e^{-Tr Z^{\dag}Z} \int d\phi
_l d \bphi _l e^{-\bphi \phi}
\label{nu1}
\ee
Since the $\phi$ integration is trivial, eq.(\ref{nu1}) defines essentially a complex
random matrix model with probability distribution $e^{-Tr Z^{\dag}Z}$. A detailed
analysis of this model has been given in \cite{ginibre}.

$Z$ is a complex matrix that can be diagonalized as, (nondiagonalizable
matrices form a set of measure zero)
\be
Z = X E X^{-1}
\label{diagonal}
\ee
where $E$ is diagonal with $E_{ii} = z_i$.
Integration over the nondiagonal part of $Z$ in (\ref{nu1}) gives the following
result \cite{ginibre}
\be
\la 1 \vert 1 \ra = \int \prod_{i=1}^{N} dz_i d\bz _ie^{-|z_i|^2} \prod_{i<j}
|z_i-z_j|^2
\label{prob1}
\ee
up to normalization factors. One can immediatelly recognize the integrand of
(\ref{prob1}) as the probability distribution corresponding to the $\nu =1$ Laughlin
wavefunction.
In A-representation we have
identified the phase space coordinates of the fermions with the eigenvalues of the
matrix
$Z$ in (\ref{coherent}).

We would like now to extend this approach in the case $p \ne 0$ in the presence of
the extra
$\Psi$ degrees of freedom. This can be viewed as a generalized complex random matrix
model. (An attempt for a random matrix formulation of the Laughlin theory of
the fractional QHE has been proposed by Callaway \cite{callaway}.) As shown in
\cite{ginibre} any complex regular
$N
\times N$ matrix
$X$ can be expressed in one and only one way as
\be
X = UYV
\label{param}
\ee
where $U$ is a unitary matrix, $Y$ is a triangular matrix with $Y_{ij}=0$ for $i>j$ and
$Y_{ii}=1$ and $V$ is a diagonal matrix with real positive diagonal elements. Using
this particular parametrization we find
\beq
&&<2p+1|Z,\phi>  = ~ [\epsilon ^{i_1...i_N} \phi_{i_1} (Z \phi)_{i_2} (Z^2
\phi)_{i_2}...( Z^{N-1}
\phi)_{i_N} ]^{2p} e^{-\12 \Tr Z^{\dag}Z} e^{-\12
\bphi\phi} \nonumber \\
&& =  [\epsilon ^{i_1...i_N} \phi_{i_1} (UYEY^{-1}U^{-1} \phi)_{i_2} (UY E^2
Y^{-1}U^{-1}
\phi)_{i_2}...( UY E^{N-1} Y^{-1}U^{-1}
\phi)_{i_N} ]^{2p} e^{-\12 \Tr Z^{\dag}Z} e^{-\12
\bphi\phi} \nonumber \\
&& = [\det (UY)]^{2p} \prod_{i<j}(z_i-z_j)^{2p} \prod_{i=1}^N
\xi_i ^{2p} e^{-\12 \Tr Z^{\dag}Z} e^{-\12 \bphi\phi}
\label {coh2}
\eeq
where $\xi = (UY)^{-1} \phi$.

Further using (\ref{diagonal}) and (\ref{param})  we have
\be
Tr Z^{\dag} Z = Tr (\bar{E} H E H^{-1})
\label{expZ}
\ee
where $H=Y^{\dag}Y$ and $\rm{det}~H=1$ ($\rm{det}~Y=1$).

In performing the integration over $\phi$'s it is convenient to change variables from
$\phi$ to $\xi$. Since $U$ is unitary and ${\rm det} Y=1$
\be
\prod _{l=1}^N d \bphi_l ~ d\phi_l ~e^{-\bphi \phi} = \prod _{l=1}^N d \bxi_l~
d \xi_l ~e^{-\bxi H \xi}
\label{psi}
\ee

Further following \cite{ginibre}, one can show that
\beq
dZ_{ij} dZ^*_{ij} & = & \prod_i dz_i d\bz_i \prod_{i < j} \vert z_i - z_j \vert ^4
\prod_{i\ne j} dH_{ij} \nonumber \\
& \equiv & d \mu (z, H) \prod_{i < j} \vert z_i - z_j \vert ^4
\label{Z}
\eeq
upto normalization factors.

Putting everything together we find
\be
\la 2p+1 \vert 2p+1 \ra = \int  d \mu (z, H) \prod_{i<j}\vert z_i - z_j \vert ^{4p+4}
   e^{- \Tr (\bar{E} H E H^{-1})}
  d \bxi_i
d \xi_i \prod_{i=1}^{N} (\bar{\xi}_i \xi_i)^{2p} e^{-\xi^{\dag} H \xi}
\label{full}
\ee

Integration over $\xi$ and $H$ would produce a quantity that depends only on
$z_i,~\bz_i$. It is clear
from the above expression that in the absence of the $\xi$ integration, the
integration over the nondiagonal elements of $Z$ would produce a probability
distribution similar to the corresponding Laughlin one. Since however
$\xi$'s couple to
$H$ and $H$ in turn couples to $z_i$ and $\bz_i$, this integration will not
necessarily produce a probability distribution which agrees with the Laughlin one. The
integration over $\xi$ and $H$ is a lot more involved now and it is hard to extract a
closed expression for arbitrary
$N$. However, as we shall see one can find general features of the
probability distribution, which do not agree with what one would expect from the
Laughlin distribution.

We first perform the $\xi$-integration by introducing a source term as follows
\beq
&& \int d \xi_i d \bar{\xi}_i \prod_{i=1}^N (\xi_i \bar{\xi}_i)^{2p} e^{-\bar{\xi} H
\xi}
\nonumber \\
&=& \prod_{i=1}^N \bigl({\del \over \del J_i} {\del \over \del \bJ_i} \bigr)^{2p} \int d
\xi_i
d \bar{\xi}_i e^{-\bxi H \xi + \bJ \xi + J \bxi} |_{J,\bJ=0} \nonumber \\
&=& \prod_{i=1}^N \bigl({\del \over \del J_i} {\del \over \del \bJ_i} \bigr)^{2p} e^{\bJ
H^{-1} J} |_{J,\bJ=0}
\label{source}
\eeq
Eq. (\ref{full}) can now be written as
\be
\la 2p+1 \vert 2p+1 \ra = \int \prod_i dz_i d\bz_i \prod_{i < j} \vert z_i - z_j \vert
^{4p+4} \prod_{i=1}^N \bigl({\del \over \del J_i} {\del \over \del \bJ_i} \bigr)^{2p}
\int \prod_{i\ne j} dH_{ij} e^{ - \Tr (\bar{E} H E H^{-1}) +\bJ H^{-1} J} |_{J,\bJ=0}
\label{source1}
\ee
We are now left with the $H$ integration
\be
I=\int \prod_{1 \le i \ne j \le N} dH_{ij} e^{- \Tr (\bar{E} H E H^{-1})}
e^{\bJ H^{-1} J}
\label{Hint}
\ee
To do this we follow an iterative procedure as in \cite{ginibre}. At
each step of the iteration we integrate over the variables of the last
row and column of $H$ and thus decrease by one the size of the matrix. In
the absence of the source terms, the structure of the reduced matrix
remains the same and this produces a simple recursion formula, which
eventually leads to (\ref{prob1}). This is not quite the case when there
are source terms, as a result there is no simple recursion formula.

The iteration procedure is defined as follows: Let $H',~E'$, etc., be the relevant
matrices of order $n$ and $H,~E$, be those obtained from $H',~E'$ by removing the last
row and last column. Greek (Latin) indices run from 1 to $n$ ($n-1$). Let $\D '_{\a \b}$
be the cofactor of $H '_{\a\b}$ in $H'$ and $\D_{ij}$, the cofactor of $H_{ij}$ in $H$.
Let $g_i=H'_{in}$. Because $\det H'=\det H=1$, the following relations are true
\cite{ginibre},
\beq
&& \D'_{\a\b}= H_{\b\a}^{'-1},~~~~~~~~~~\D_{ij} = H_{ji}^{-1} \nonumber \\
&& H'_{nn}=1 + \sum_{i,j} g_i^{*} g_{j} \D_{ji} \nonumber \\
&& \D'_{in} = - \sum_{l} \D_{il} g^{*} _l \label{relations} \\
&& \D'_{ij} = H'_{nn} \D_{ij} - \sum_{l,k} g^{*}_{k}g_{l} \D_{ij}^{lk} \nonumber \\
&& \D_{ij}^{lk} = \D_{ij} \D_{lk} - \D_{ik} \D_{lj} \nonumber
\eeq
Let
\be
\Phi_{n}^{J} = \Tr \bigl( \bar{E}' H' E'H^{'-1} \bigr) - \bJ ' H^{'-1} J' \equiv
\Phi_{n}^{0} - \bJ ' H^{'-1} J'
\label{phin}
\ee
Using (\ref{relations}) we find
\beq
\Phi_{N}^{J}  = && \vert z_N \vert ^2 + \Phi_{N-1}^{J} -J_N \bJ_{N} + \la g \vert
H^{-1} (\bar{E} -\bz_{N}) H ( E-z_N)H^{-1} \vert g \ra \nonumber \\
&& - \la g \vert  H^{-1} J \bJ H^{-1} \vert g \ra + g_{l}^{*} (H^{-1}J)_{l} \bJ_{N} +
J_{N} (\bJ H^{-1})_l g_l  \label {iteration}
\eeq
Integration over $g$'s produces the following result
\be
I \sim \int \prod_{1 \le i \ne j \le N-1} dH_{ij}  e^{-\vert z_N \vert ^2}
e^{-\Phi_{N-1}^{J}} e^{J_N
\bJ _N} {{e^{\left( \sum \bJ_i X_{ij}^{-1} J_{j} \right) J_N \bJ_N}} \over {\det \bigl(
X_{ij} -J_i
\bJ _{j}\bigr)}}
\label{first}
\ee
where
\be
X_{ij} = \bigl( H^{-1} (\bar{E}-\bz_N) H (E-z_N) H^{-1}\bigr)_{ij} - J_i \bJ_j =
X_{ij}^{0} -J_i \bJ_j
\label{X}
\ee
The expression (\ref{first}) can be further simplified using the following relations
\be
\bJ_i \bigl(X^0 -J \bJ\bigr)^{-1}_{ij} J_j = \bJ _i \bigl( (X^0)^{-1} + (X^0)^{-1} J \bJ
(X^0)^{-1} +... \bigr) _{ij} J_j = {{\bJ _i (X^0)^{-1}_{ij} J_j} \over {1-\bJ _i
(X^0)^{-1}_{ij} J_j}}
\label{s1}
\ee
Further
\beq
&&\det\bigl( X^0 - J \bJ \bigr) = e^{\Tr \log \bigl( X^0 - J \bJ \bigr)}=
e^{\Tr\bigl[ \log X^0 + \log(1-(X^0)^{-1} J \bJ) \bigr] }\nonumber \\
&& = \det (X^0)~ e^{\log \bJ (X^0)^{-1} J} = \prod_{i<N} \vert z_i-z_N \vert ^2
\sum_{1 \le l,m \le N-1} \bJ_l (X^0)^{-1}_{lm} J_m
\label{s2}
\eeq
where we have used that
\be
 \det X^0 = \det \bigl( H^{-1} (\bar{E}-\bz_N) H (E-z_N)
H^{-1}\bigr) = \prod _{i< N} \vert z_i-z_N \vert ^2
\ee
Substituting (\ref{s1}, \ref{s2}) in
(\ref{first}) we find
\be
I \sim {e^{- \vert z_N \vert ^2} \over
{\prod_{i<N} \vert z_i-z_N \vert ^2}} \int \prod_{1 \le i \ne j \le N-1} dH_{ij}
{e^{-\Phi_{N-1}^{J}}
\over
 (1-M)} e^{{J_N
\bJ_N} \over {1-M }}
\label{I}
\ee
where
\be
M= \sum_{1 \le i,j \le N-1} \bJ_i (X^0)^{-1}_{ij} J_j = \sum_{1 \le i,j \le N-1} {\bJ _i
\over {(z_i - z_N)}} H^{-1} _{ij} {J_j \over {(\bz _j -\bz _N)}}
\label{M}
\ee
Already this result can be used to evaluate (\ref{full}) for the simple case of a
$2 \times 2$ matrix
model. Although this is a rather trivial case, it is worth presenting it, since it
highlights properties of the probability distribution which are not in agreement with
the Laughlin one.

To simplify the calculation let us further choose $p=1$. For the $N=2$ case after
the first iteration we find from (\ref{I})
\be
I \sim {{e^{- \vert z_1 \vert ^2} e^{- \vert z_2 \vert ^2}}\over
{\vert z_1-z_2 \vert ^2}} {e^{J_1 \bJ_1} \over {1-M}}
 e^{{J_2
\bJ_2} \over {1-M }}
\ee
where $M= 1-{{J_1 \bJ_1} \over {\vert z_1-z_2 \vert ^2}}$.The functional diferentiation
with respect to $J$'s, eq. (\ref{source}), gives
\beq
&& \left( {\del \over {\del J_1}}{\del \over {\del \bJ_1}}\right) ^2 \left( {\del \over
{\del J_2}}{\del \over {\del \bJ_2}}\right) ^2~I \vert _{J,\bJ =0} \nonumber \\
&& = 2 {{e^{- \vert z_1 \vert ^2} e^{- \vert z_2 \vert ^2}}\over
{\vert z_1-z_2 \vert ^2}} \left( {\del \over {\del J_1}}{\del \over {\del \bJ_1}}\right)
^2 {e^{J_1 \bJ_1} \over {\left(1 - {{J_1 \bJ_1} \over {\vert z_1-z_2 \vert
^2}}\right)^3}}
\vert_{J_1,
\bJ_1 =0} \nonumber \\
&& = {{e^{- \vert z_1 \vert ^2} e^{- \vert z_2 \vert ^2}}\over
{\vert z_1-z_2 \vert ^2}} \left[ 4 + {24 \over {\vert z_1-z_2 \vert ^2}} + {48
\over {\vert z_1-z_2 \vert ^4}}\right]
\label{I2}
\eeq
Using (\ref{full}), (\ref{source}) and (\ref{I2}) we find
\be
\la 3 \vert 3 \ra \sim \int dz_1  d\bz _1 dz_2 d\bz _2 e^{-\vert z_1 \vert^2-\vert z_2
\vert ^2} \vert z_1 - z_2 \vert ^6 ~\left[ 1 + {6 \over {\vert z_1-z_2 \vert ^2}} + {12
\over {\vert z_1-z_2 \vert ^4}}\right]
\label{nu3}
\ee
The first term corresponds to the probability distribution for the $\nu = 1/3$ ground
state Laughlin wavefunction. There are extra terms though, which are dominant at
short distances as $z_1 \rightarrow z_2$. In this simple case we find that the
distribution emerging from the matrix model has a long distance behavior similar to the
corresponding Laughlin one, but its short distance behavior is quite different. We
shall now argue that
this behavior prevails
for any
$N$.

It is clear from (\ref{I}) that the first step of the iteration produces an expression
which is not quite similar to the original one if
$J's
\ne 0$. As a result the integration over $H'_{in}$ at each subsequent step of the
iteration becomes quite involved. Although it is very hard to derive an exact
expression for $I$ as a function of $z_i$'s, it is quite straightforward to explore its
dependence on $\vert z_{N-1} - z_N \vert$. This is sufficient for example to get
information about the short and long distance behavior of the probability distribution
as  $\vert z_{N-1} - z_N \vert << 1$ and $\vert z_{N-1} - z_N \vert >> 1 $ respectively.

The dependence of $I$ on $\vert z_{N-1} - z_N \vert$ comes first from the overall
factor $1 / \prod_{i<N} \vert z_i - z_N \vert ^2$ and second from the factor $M$, see
eq. (\ref{I}), (\ref{M}). Expanding $M$ we find
\beq
M &=&{{\bJ _{N-1} \bJ_{N-1}}
\over {\vert z_{N-1} - z_N \vert ^2 }} + {\bJ _{N-1}
\over {(z_{N-1} - z_N)}} \sum_{j < N-1} H^{-1} _{N-1,j} {J_j \over {(\bz _j -\bz
_N)}} + \sum_{i <N-1}{\bJ_i \over {(z_i -z
_N)}} H^{-1} _{i,N-1} {J _{N-1}
\over {(\bz_{N-1} - \bz_N)}} \nonumber \\
&& + \sum_{1 \le i,j \le N-2} {\bJ _i
\over {(z_i - z_N)}} H^{-1} _{ij} {J_j \over {(\bz _j -\bz _N)}}
\eeq
It is clear now that in order to explicitly demonstrate the $\vert z_N - z_{N-1} \vert
$ dependence we need to evaluate the functional derivatives with respect to $J_N$ and
$J_{N-1}$, as in (\ref{source1}). We find
\beq
&&\prod _{i}\left( {\del \over {\del J_i}}{\del \over {\del \bJ_i}}\right) ^2 ~I \vert
_{J, \bJ = 0} =  {{e^{- \sum_i \vert z_i \vert ^2}}\over
{\prod_{i<N} \vert z_i-z_N \vert ^2}} ~ \bigl( A + {B \over {z_{N-1}-z_N}} + {B^* \over
{\bz_{N-1}-\bz_N}}
\nonumber \\
&& + {C \over {\vert z_{N-1}-z_N \vert ^2}} + {D \over {( z_{N-1}-z_N)
^2}} + {D^* \over {( \bz_{N-1}-\bz_N)  ^2}} \nonumber \\
&&+ {1 \over {\vert z_{N-1}-z_N
\vert ^2}} \left({E \over {z_{N-1}-z_N}} + {\bar{E} \over
{\bz_{N-1}-\bz_N}} \right) + {F \over {\vert z_{N-1}-z_N \vert ^4}} \bigr)
\eeq
where $A,~B,~C,~D,~E,~F$ are in general functions of $z$'s containing factors of the
form $(z_i-z_N),~ (z_i- z_{N-1}),~ (z_i-z_j)$ and their complex conjugates, where $i \le
N-2$. Substituting this expression in (\ref{source1}) we find a probability distribution
which is dominated by a term proportional to $\vert z_{N}-z_{N-1} \vert ^6$ when $\vert
z_{N}-z_{N-1}
\vert
>> 1$, similar to the $\nu = 1/3$ Laughlin distribution. However when $\vert
z_{N}-z_{N-1} \vert << 1$ the matrix distribution is dominated by a term proportonal to
$\vert z_{N}-z_{N-1} \vert ^2$. The long and short distance behavior for any $N$ is the
same as the one found in the simple $N=2$ case, eq.(\ref{nu3}).

It is straightforward now to see that similar results can be derived for any $p \ne 0$.
In particular the long distance behavior of the probability distribution is that of the
$\nu = {1 \over {2p+1}}$ Laughlin distribution while the short distance behavior is that
of a $\nu =1$ Laughlin distribution.

Further the probability distribution cannot
be factorized as
$\Psi^*
\Psi$ where
$\Psi$ is the corresponding many-particle wavefunction, which is both antisymmetric
and holomorphic in $z$'s. This indicates that the identification of the eigenvalues
of the matrix
$Z$ with the actual holomorphic coordinates of fermions may not be appropriate.

\noindent
$\underline{X-{\rm representation}}$

Going to an $X$-representation
first we derive one-dimensional fermionic wavefunctions, where the coordinates have
been identified with the eigenvalues of the matrix $X_1$, then
transform the wavefunctions to the coherent state
representation.

We define the state $\vert X, \phi \ra$ such that
\be
\hat{X_1} \vert X, \phi \ra = X \vert X, \phi \ra ~~~~~~~~~~~~~~\Psi
\vert X , \phi
\ra = \phi \vert X , \phi \ra
\label{xrep}
\ee
Using (\ref{xrep}) we find
\be
\la 2p+1 \vert X,\phi \ra = \bigl[ \epsilon ^{i_1...i_N} \phi_{i_1} (A
\phi)_{i_2}...(A^{N-1}\phi)_{i_N} \bigr] ^{2p} ~\la 0 \vert X, \phi \ra
\label{px}
\ee
where
\beq
\la 0 \vert X, \phi \ra & = & e^{-\Tr {1 \over 2} B X^2} e^{-\12 \bphi \phi} \nonumber
\\ A_{ij} & = & \sqrt{{B \over 2}} \bigl( X_{ij} - {1 \over B} {\del \over {\del
X_{ji}}}
\bigr)
\label {A}
\eeq
Since (\ref{px}) is completely antisymmetric in the $i_n$-indices, the differential
operator ${\del \over {\del X_{ji}}}$ produces a nonzero contribution only if it acts
on the ground state wavefunction $\la 0 \vert X, \phi \ra$. We then find
\be
\la 2p+1 \vert X, \phi \ra  =  (\sqrt{2B})^{pN(N-1)} \bigl[ \epsilon ^{i_1...i_N}
\phi_{i_1} (X \phi)_{i_2}...(X^{N-1} \phi)_{i_N} \bigr] ^{2p} e^{-\Tr {1 \over 2}B X^2}
e^{-\12
\bphi \phi}
\label{X}
\ee
$X$ being a hermitian matrix, it can be diagonalized by a unitary transformation
\be
X= U x U^{-1} ~~~~~~~~~~~~~~x_{ij}= x_i \d _{ij}
\label{diag}
\ee
Using this in (\ref{X}) we find
\beq
\la 2p+1 \vert X, \phi \ra  & = & (\sqrt{2B})^{pN(N-1)} \bigl[ \epsilon ^{i_1...i_N}
\phi_{i_1} (U x U^{-1} \phi)_{i_2}...(U x^{N-1} U^{-1} \phi)_{i_N} \bigr] ^{2p} ~
e^{-\Tr {1
\over 2}B X^2} e^{-\12 \bphi \phi} \nonumber \\
& = & (\sqrt{2B})^{pN(N-1)} \left[\det U \right]^{2p} \prod _{i<j} \bigl( x_i -x_j \bigr)
^{2p}
\prod _{i=1}^{N}
\bigl( U^{-1} \phi \bigr)^{2p}_i ~ e^{-\Tr {1
\over 2}B X^2} e^{-\12 \bphi \phi}
\label{xi}
\eeq
Using (\ref{xi}) we can express the scalar product of the $\vert 2p+1 \ra$ state as
\beq
\la 2p+1 \vert 2p+1 \ra & = & \int \bigl[ dX_{ij} \bigr] d \bphi _l d \phi _l \la 2p+1
\vert X, \phi \ra \la X, \phi \vert 2p+1 \ra \nonumber \\
& \sim & \int \bigl[ dX_{ij} \bigr] \prod _{i<j} \bigl( x_i
-x_j\bigr) ^{4p} e^{-\Tr {B X^2}} d
\bphi _l d \phi _l \prod_{i}
\bigl[ (U^{-1} \phi )_i (\bphi U)_i \bigr]^{2p} e^{-\bphi \phi}
\label{scalarx}
\eeq
Since $U$ is a unitary matrix the integration over $\phi$'s completely decouples
(unlike the case in eq.(\ref{full})). Further \cite{brezin}
\be
\bigl[ dX_{ij} \bigr] = dx_i \prod _{i<j} \bigl( x_i
-x_j\bigr) ^2 [dU]
\label {measur}
\ee
where $[dU]$ is the Haar measure. Integration over the nondiagonal elements of $X$ gives
\be
\la 2p+1 \vert 2p+1 \ra  \sim \int dx_i \prod_{i<j} (x_i - x_j)^{4p+2} e^{-B\sum_{i}
x_i^2}
\ee
This is the probability distribution for the one-dimensional Calogero ground state
wavefunction $\prod_{i<j} (x_i - x_j)^{2p+1} e^{-B\sum_{i}
x_i^2 /2}$. This is not surprising; it was already indicated in \cite{poly1, poly2} that
the Chern-Simons matrix model is equivalent to the Calogero model.

Let us now use a coherent state representation for the Calogero wavefunction $\la
x_1,...,x_N \vert \Psi \ra = \prod_{i<j} (x_i -x_j)^{2p+1} e^{-B\sum_i x_i^2 /2}$. The
coherent state representation of any wavefunction $\la x \vert \Psi \ra$ can be
written as
\be
\la z \vert \Psi \ra = \int dx \la z \vert x \ra \la x \vert \Psi \ra
\label {zx}
\ee
where $\hat{z} \vert z \ra = z \vert z \ra $ and $\hat{z} = \sqrt{B \over 2} (\hat{x} +
i \hat{y}),~~~[\hat{x},~ \hat{y}] = {i \over B}$.
Using
\be
\la z \vert x \ra = e^{-{B \over 2} x^2} e^{\sqrt{2B} \bz x} e^{-{\bz^2 /2}}
e^{{-\vert z \vert ^2}/ 2}
\ee
we find
\be
\la z \vert \Psi \ra = \int dx e^{-B (x-{\bz \over {\sqrt{2B}}})^2} f(x)
\label{int}
\ee
where $f(x)$ is given by $\la x \vert \Psi \ra = f(x) e^{-B x^2 /2}$.
In evaluating (\ref{int}) we expand $f(x)$ around $\bz / \sqrt{2B}$.
\beq
\la z \vert \Psi \ra & = & e^{-{{\vert z \vert ^2}\over 2}} \sum _{k=0}^{\infty} \int dx
e^{-B (x-{\bz \over {\sqrt{2B}}})^2}{1 \over {(2k)!}} (x-{\bz \over
{\sqrt{2B}}})^{2k} ~{{\del ^{2k} f} \over { \del x^{2k}}} \vert _{x = \bz /\sqrt{2B}}
\nonumber \\
& = & e^{-{{\vert z \vert ^2}\over 2}} \sum_{k=0}^{\infty}{{\Gamma (k + {1 \over 2})}
\over {(2k)!~ B^{k + {1
\over 2}}}} ~{{\del ^{2k} f} \over { \del x^{2k}}} \vert _{x = \bz /\sqrt{2B}} \nonumber
\\ & = & \sqrt{\pi \over B} e^{-{{\vert z \vert ^2}\over 2}} \bigl[ e^{{1 \over {4B}}
{\del^2 \over {\del x ^2}}} f(x) \bigr] _{x = \bz /\sqrt{2B}}
\eeq
The coherent state representation of the many-body Calogero wavefunction is thus
\be
\la z_1,...,z_N \vert \Psi \ra \sim e^{-\sum_i \vert z_i \vert^2 /2} ~\bigl[ e^{{1
\over {4B}}
\sum_i
{\del^2 \over {\del x _i^2}}} \prod _{i<j} (x_i -x_j)^{2p+1} \bigr] _{x_i = \bz_i
/\sqrt{2B}}
\label{zC}
\ee
For $p=0$ (\ref{zC}) gives the $\nu =1$ Laughlin state. For $p \ne 0$ the coherent
representation of the Calogero state has a long distance behavior similar to the $\nu =
{1 \over {2p+1}}$ Laughlin state, but a different short distance behavior. This
connection between the one-dimensional Calogero wavefunction and the Laughlin states
has already been noted in \cite{iso}.

\vskip .3in
{\bf 5. Summary and discussion}

In an attempt to clarify the exact correspondence between the Chern-Simons matrix model
introduced by Polychronakos and the fractional QHE at filling fraction $\nu = 1/m$, as
described by Laughlin wavefunctions, we have derived the matrix model wavefunctions using a
coherent state representation. We have presented two different coherent state
representations, each one implementing a different choice for the phase space coordinates
of the underlying one-dimensional fermionic system.

In the $A$-representation, the eigenvalues of the matrix $A$ are identified with the phase
space coordinates $z$ of the fermions, while in the $X$-representation, the eigenvalues of
the matrix $X_1$ are identified with the one-dimensional coordinates $x$ of the fermions.

Both choices give identical results when $p=0$, or equivalently $\theta =0$ in
(\ref{action}). The corresponding wavefunction is identical to the $\nu=1$ Laughlin
wavefunction.

For $p \ne 0$ the two representations give different results.
Although the explicit expressions for the probability distributions are different, they
share some common features. They both have the same long and short distance behavior.
Comparing them to the corresponding
$\nu = {1
\over {2p+1}}$ Laughlin distributions, we find that it is only the long distance
behavior which is in agreement. The short distance behavior does not agree with the
Laughlin one.

A noticeable difference between the two representations is that the $X$-representation
leads to a holomorphic wavefunction with antisymmetric properties, while this is not
possible for the $A$-representation. 

As we mentioned earlier there is an ambiguity in introducing electron coordinates in the
matrix model. This has to do with the choice of coherent state representation. In this
paper we analyzed two particular coherent state representations, which seem to be the
natural choices from the matrix model point of view. In both cases we find that the
emerging wavefunctions do not quite agree with the Laughlin one. Although this by itself
does not prove that the original matrix model is not equivalent to the Laughlin theory for
the $\nu =1/m$ fractional QHE, it makes the precise correspondence between the two models
less transparent. One can argue that there may exist another coherent state
representation, corresponding to another choice for the electron coordinates, such
that there is agreement with the Laughlin wavefunction. However, for the matrix
model to be truly useful in the context of QHE, this new coordinate choice should be easily
identifiable.

\vskip .3in
{\bf Acknowledgements}

We would like to thank A.P. Polychronakos for many useful discussions on the matrix
model and its relation to the Calogero model. Discussions with V.P. Nair are also
acknowledged. This work was supported in part by the NSF grant PHY-9970724 and a PSC-31
CUNY award.

\end{document}